\documentclass[twocolumn,showpacs,preprintnumbers,amsmath,amssymb]{revtex4}
\usepackage{graphicx}
\usepackage{color}
\def\beq{\begin{equation}}
\def\eeq{\end{equation}}
\def\beqa{\begin{eqnarray}}
\def\eeqa{\end{eqnarray}}

\begin{document}
\title{Beam-dust interactions in an e$^+$e$^-$ collider}
\author{
Kazuhito~Ohmi$^{\mbox{\scriptsize 1,2,3}}$, Hitoshi~Fukuma$^{\mbox{\scriptsize 1}}$
and Shinji~Terui$^{\mbox{\scriptsize 1}}$ \\
$^{\mbox{\scriptsize 1}}$KEK, High Energy Accelerator Organization, 
Tsukuba, Ibaraki 305-0801, Japan \\
$^{\mbox{\scriptsize 2}}$IHEP, Institute of High Energy Physics,
Yuquan Road Beijing 100049, P.R.China\\
$^{\mbox{\scriptsize 3}}$University of Science and Technology of China, 
Hefei,Anhui, 230026, P.R.China\\
}

%
\begin{abstract}  
In recent electron-positron colliders designed and operated with very low emittance and high current, 
the energy density of the beam has increased significantly compared to earlier designs. 
Under these conditions, interactions between the beam and residual materials within 
the beam pipe exert considerable mutual influence. 
The residual materials are heated by the beam, leading to evaporation, charging, and possible fission, 
eventually resulting in a plasma state. 
Conversely, the beam  undergoes emittance growth due to electromagnetic interactions 
with the resulting plasma. 
We investigate the effects of such dust-induced plasma on the beam through numerical simulations 
of these processes.
\end{abstract}
%
\maketitle

\section{Introduction}
The effects of dust on beams were initially observed in electron storage rings 
used as synchrotron radiation sources \cite{FZPAC95,Heifets,DustBEPC,tanimotoPRSTAB,Novok}. 
In recent years, dust-related effects have also been detected in proton beams at the LHC, 
manifested as unidentified falling object (UFO) events \cite{LindLHC,ALechLHC,BelangerLHC}. 
More recently, sudden beam losses (SBL) have been observed primarily in the positron ring (LER) of 
SuperKEKB, which are suspected to be caused by dust. 

The beam losses in SuperKEKB occur within a very short time span of a few turns ($10\sim 50 \mu$s). 
In some instances, such losses have induced quenches in superconducting magnets near 
the collision point. This phenomenon represents one of the most critical issues 
in achieving higher beam currents.
The characteristics of the beam loss can be summarized as follows:
\begin{itemize}
\item In most cases, it occurs under high beam current conditions of over 500~mA.
\item The beam loss begins abruptly without any precursor of dipole oscillation and leads to an abort 
    within one to several turns.
\item It occurs independently of beam-beam collisions.
\item It is more frequent in the 4 GeV positron ring (LER) but also occurs in the 7 GeV electron ring (HER).
\end{itemize}

In the collider, collimators are positioned as closely as possible to protect the detector 
and minimize background noise. For SuperKEKB, with a vertical emittance of
$\varepsilon_y=5\times 10^{-11}$~m \cite{ohnishi_eefact25}, 
the physical aperture is $2J_{y,max}=0.8\times 10^{-7}$~m \cite{suetsugu}, 
corresponding to $40~\sigma_y$. 
This implies that positrons in the beam can exceed this amplitude within 
a few turns even in the absence of coherent dipole oscillations.

To investigate the cause, extensive experiments were conducted at the SuperKEKB accelerator 
\cite{terui_eefact25}. 
Beam aborts were intentionally triggered by physically striking the beam pipe in the wiggler section. 
Initial hypotheses considered copper (from the beam pipe) and 
silicon (from VACSEAL, a material commonly used for vacuum sealing repairs) 
as potential atomic gas sources. Simulations of beam-dust interactions 
based on copper and silicon dust were presented in Ref.~\cite{ohmiSBL_eefact25}.
Upon inspection of the beam pipe, particles several tens of micrometers in size were observed, 
and compositional analysis identified them as graphite. 
This graphite is thought to be organic matter that has been carbonized by irradiation with 
synchrotron radiation.

The beam size immediately after the abort was measured turn-by-turn using a gated camera 
and a streak camera. An increase in the vertical beam size was observed at the occurrence of 
SBL \cite{IkedaIBIC25}.

In this paper, we discuss the interaction between a positron beam and graphite dust by 
means of simulation.
Graphite sublimates directly from a solid to a gas at approximately 3500 K; 
no liquid phase exists under the vacuum conditions of the beam pipe.
The beam-dust interaction model was subsequently improved to include the absorption of 
ionization-generated electrons onto dust surfaces and subsequent secondary electron emission. 
Furthermore, the use of graphite, a sublimating material, introduces a qualitatively distinct 
physical regime.

Initially, simulations were performed under the assumption that vacuum bumps 
were the cause of beam aborts.
These simulations revealed that when approximately $10^{10}$~m ions 
are generated near the beam, 
the beam experiences strong defocusing, resulting in amplitudes large enough to induce beam loss.
This type of loss stems from an instantaneous mismatch in the Twiss parameters, 
implying that dipole oscillations are not necessarily observed.
This mechanism is discussed in Sec.~\ref{VacBump}.

The fundamental physical properties of dust particles under beam irradiation are examined 
in Sec.~\ref{BDpar}.
The simulation accounts for beam ionization loss in a dust, electron production, electron absorption, 
secondary electron emission, evaporation, and ionization of atomic gas.
Charged particles, including the beam and ionized atomic gases, are represented as macroparticles, 
and their motion is tracked within the electric field generated by the ensemble using 
the Particle-In-Cell (PIC) method.
Building on these physical principles, the simulation of beam-dust interactions is presented in 
Sec.\ref{BDsim}.

The SBL phenomenon in SuperKEKB is observed on the timescale of several turns. 
Conducting a multi-turn simulation involving all bunches is computationally challenging 
due to resource limitations. 
Therefore, a simplified model was used to perform a multi-turn simulation, 
as described in Sec.~\ref{MultiTurn}.


\section{Interaction between residual gas and the beam}\label{VacBump}
We begin by examining the case of a sudden pressure bump. 
Similar approaches have been employed in studies of electron cloud instability \cite{PEI,PEHT} 
and ion instability \cite{FII}, making this method a suitable initial step 
for understanding the overall phenomenon.
Assuming a pressure of $P=2$~Pa at a temperature $T=300$~K or $P=20$~Pa at $T=3000$~K
within a region of $l=1$~m, we evaluated its effect on the beam.
The gas is assumed to be composed of carbon, which has an atomic number $Z_i=6$ and a mass number $A_i=12$.
The corresponding integrated density over a length $1$~m is $\rho_G=5.3\times 10^{20}$~m$^{-2}$. 

The effect of beam collisions with residual gas molecules in the beam pipe is a fundamental 
and critical concern in accelerator design and operation, and extensive research has been conducted
on this topic. From the perspective of beam lifetime in GeV-class electron and positron beams, 
Coulomb scattering with atoms, which significantly alters the transverse momentum of beam particles,
and bremsstrahlung, which substantially changes their energy, are dominant effects and are routinely 
evaluated during the design phase.

Residual gas is neutral, so its interaction with the beam is primarily characterized by 
Coulomb scattering. The cross-section for Coulomb scattering is given by:
\beq
\sigma_{Coul}=\frac{4\pi r_e^2 Z_i(Z_i+1)}{\gamma^2\theta_c^2}
\eeq
where $\gamma$ is  the relativistic factor, $r_e$ is the classical electron radius, and
$\theta_c$ is the minimum angular spread that leads to beam loss, defined as:
\beq
\frac{1}{\theta_c^2}=
\frac{\beta_x}{4J_{x,max}}+\frac{\beta_y}{4J_{y,max}}
\eeq
Here, \(2J_{x,\text{max}}\) and \(2J_{y,\text{max}}\) represent the transverse acceptance 
in terms of action variables, and \(\beta_{x,y}\) denotes the beta function at the location 
where Coulomb scattering occurs.
The beam loss rate per collision is given by
\beq
\frac{T_0}{\tau}=\frac{P l}{k_B T}\sigma_{Coul}. \label{BGlife}
\eeq
where $k_B$ is the Boltzmann constant, $\tau$ the lifetime and $T_0$ the revolution time.

In general, for high-energy electrons, energy loss due to bremsstrahlung,
which causes particles to be lost outside the energy aperture, is also significant. 
However, in the case of SuperKEKB, the transverse aperture is constrained by collimators, 
making Coulomb scattering the dominant loss mechanism~\cite{suetsugu}. 
For graphite (\(Z_i = 6\)) at \(T = 300\)~K and \(Pl = 2\)~Pa\(\cdot\)m, 
the lifetime is calculated to be 1.1 seconds (\(1.1 \times 10^5\) turns), 
independent of beam current and emittance.

Next, we consider the scenario in which the gas becomes ionized. 
Using the integrated gas density \(\rho_G = 5.3 \times 10^{20}\)~m\(^{-2}\) and
assuming an ionization cross-section of \(\sigma_{\text{ion}} = 2 \times 10^{-22}\)~m\(^{-2}\), 
the number of ions produced per bunch passage (with \(N_p = 6.25 \times 10^{10}\) particles) 
is \(\Delta n_i = 6.6 \times 10^9\), corresponding to a density increment of 
\(\Delta \rho_i = 2.0 \times 10^{17}\)~m\(^{-2}\). 
Electrons are generated simultaneously in equal numbers.
The positron beam focuses electrons and defocuses ions, while it is itself focused by 
electrons and defocused by ions. Since the spatial distributions of electrons and ions differ, 
their effects do not cancel, leading to net changes in beam dynamics.

As the bunch train traverses the region, electron and ion clouds are formed. Each bunch interacts 
with the electron or ion cloud, altering its motion. 
To simulate the generation of electron/ion clouds and their interaction with bunches, 
each bunch of length \(\sigma_z = 6\)~mm is divided into 40 slices along the direction of propagation. 
Whenever a bunch passes through the gas, macro-particles representing electrons and ions are generated. 
The electrostatic potentials of both the electron/ion clouds and each bunch slice 
are computed using the Particle-In-Cell (PIC) method, and their mutual interactions and motions are tracked.
The number of macro-particles is fixed at \(10^6\) for the positron bunch, 
while it varies between \(10^5\) and \(10^6\) for ions and electrons.

The beam emittance of SuperKEKB is \(\varepsilon_x = 4\)~nm and \(\varepsilon_y = 50\)~pm, 
with average beta functions \(\beta_x = \beta_y = 10\)~m in both planes. 
The beam sizes are \(\sigma_x = 200~\mu\)m and \(\sigma_y = 22~\mu\)m. 
The electron plasma frequency is:
\[
\omega_e = \left( \frac{N_p r_e c^2}{\sqrt{2\pi} \sigma_z \sigma_x \sigma_y} \right)^{1/2}
 = 2\pi \times 90~\text{GHz},
\]
and the number of slices per bunch (40) exceeds \(\omega_e \sigma_z / c \approx 9.4\). 
The bunches pass through the gas every 4~ns.

\begin{figure}[hbt]
    \centering
    \includegraphics[width=40mm]{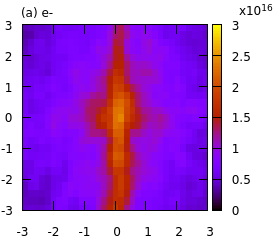}
    \includegraphics[width=40mm]{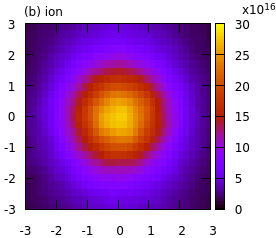}
    \includegraphics[width=40mm]{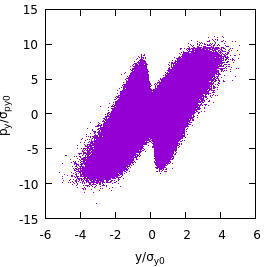}
    \caption{Distribution of electrons, ions, and beam particles after 10 bunch passages.
    Plots (a) and (b) display the electron and ion distributions, respectively.
    Plot (c) displays the vertical phase space distribution of the beam particles 
    in the 10th bunch. 
    The  horizontal and vertical axes represent 
    $x/\sigma_x$ and $y/\sigma_y$, respectively.
    The color scale in (a) and (b) represents density in units of m$^{-2}$.}
    \label{eibdist}
\end{figure}

Figure \ref{eibdist} shows the distributions of electrons, ions, and beam particles 
after 10 bunch passages.
Plots (a) and (b) depict the electron and ion distributions, respectively.
The horizontal and vertical axes represent $x/\sigma_x$ and $y/\sigma_y$, where $\sigma_{x,y}$
denotes the beam size.
Plot (c) shows the vertical phase space distribution of the beam particles in the 10th bunch.

Near the center, the bunch is focused due to the electron cloud.
Overall, however, the bunch experiences defocusing caused by the ion cloud.
From the slope of the phase space distribution, it can be inferred that the focusing component, 
though narrow in range, is several times stronger than the defocusing component. 
Plot (a) shows the distribution after bunch passage; however, it is reasonable to assume that 
during bunch passage, the electron distribution is more highly localized than that of ions.
The defocusing strength is estimated as $K\beta_y\approx 5$.
This value is relatively large, corresponding to a tune shift of approximately 0.4, 
but it is not sufficient to cause sudden beam loss.

\begin{figure*}[hbt]
    \centering
    \includegraphics[width=140mm]{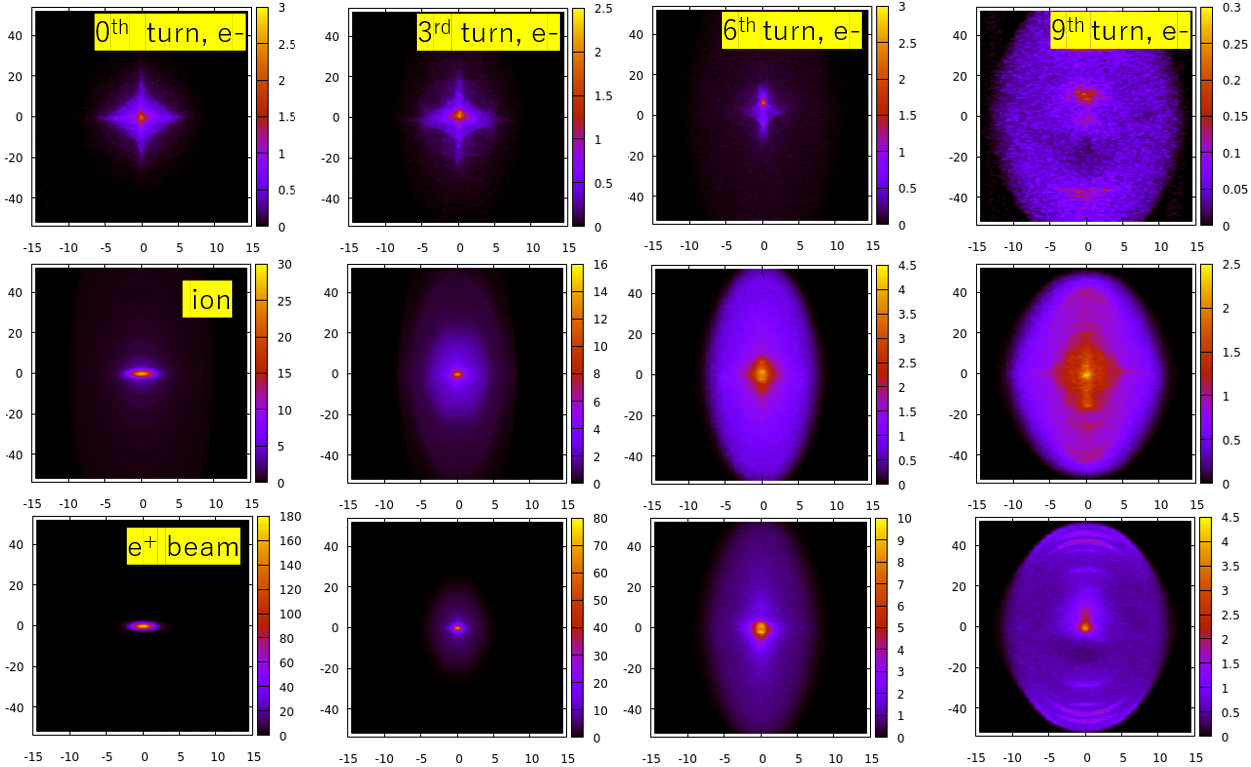}
    \caption{Electron, ion and beam distributions during the interaction with the 10th bunch after
    0, 3, 6 and 9 turns. The color scale represents density in units of $10^{16}$~m$^{-2}$.}
    \label{eibdist2}
\end{figure*}

We investigate the multi-turn effects of beam interactions with residual gas.
The ion distribution saturates around the 10th bunch passage.
Subsequently, the bunch interacts with this saturated distribution over multiple turns.
Figure~\ref{eibdist2} shows the electron, ion, and beam distributions for the 10th bunch 
after several turns (as indicated in the figure).
The beam distribution expands to  $\approx 50\sigma_y$ by the 4th turn.
This size is sufficient to cause beam loss.
 
A similar simulation has been conducted for copper (mass number $A_i=64$) in Ref.\cite{ohmiSBL_eefact25}.
For graphite with mass number $A_i=12$, the phase space distribution of beam particles exhibits 
a relatively clear focusing component, as observed in Fig.\ref{eibdist}. 
In the beam distribution shown in Fig.\ref{eibdist2}, the core portion remains more distinct for 
graphite compared to copper.


\section{Interaction between dust and the beam}\label{BDpar}
In this section, we review the formalisms for beam-dust interactions. Based on these, 
we will simulate the interaction between the beam and plasma-state dust in the next section.
Before proceeding, we evaluate the lifetime due to Coulomb scattering with the dust particle.
The loss rate per collision is given by
\beq
\frac{T_0}{\tau}=\frac{4N_A l}{3A_i/\rho}I_G\sigma_{Coul}. \label{BDlife}
\eeq
where $N_A=6.02\times 10^{23}$ is the Avogadro constant and $\rho$
is the mass density. $I_G$ is a coefficient that 
represents the overlap between the beam and the dust as
\beq
I_G=\frac{1}{2\pi\sigma_x\sigma_y}\int_{x^2+y^2<a^2}e^{-x^2/2\sigma_x^2-y^2/2\sigma_y^2}dxdy
\eeq
where $x$ and $y$ denote transverse coordinates, and  
$\sigma_x$ and $\sigma_y$ are the RMS beam sizes.
For graphite (C, $Z_i=6, A_i=12$, $\rho=1.8$~g/cm$^{-3}$) dust particle with $a=50~\mu$~m, the loss rate is calculated to be 0.06, 
corresponding to a beam lifetime of 35 turns.

We now examine the ionization of dust by the beam and the subsequent interaction between 
the charged dust, electron-ion clouds and the beam. The behavior of dust under beam interaction 
has been studied in Refs. \cite{Heifets,ACCHandBook}. 
However, modern beams exhibit significantly smaller emittance and higher intensity than those 
in earlier studies, necessitating investigation of rapid dynamics in beam-dust interactions.

The energy deposited in the dust per bunch passage is:
\beq
\Delta E_T=\Delta E_Q \log
\left(\frac{\varepsilon_{max}}{\varepsilon_{min}}\right)
\label{dET}
\eeq
where
\beq
\Delta E_Q=2\pi r_e^2 m_ec^2 N_A \frac{Z_i}{A_i}\frac{4 a}{3}\rho. \label{EQ}
\eeq
Here $\varepsilon_{max}$ and $\varepsilon_{min}$ is the maximum and minimum energy losses of 
recoiled (ionized) electrons in the dust, respectively.
The logarithmic term $\log(\varepsilon_{max}/\varepsilon_{min})$ is referred to as the stopping factor.
They are given by
\[\varepsilon_{max}=\frac{Qe^2}{a}+\varepsilon(a)+\varepsilon_{min}\]
where $Q$ is the dust charge in units of electron charge.
$\varepsilon_{min}=(\hbar\omega_p)^2/(2m_ec^2)$ is adopted from Ref.\cite{Heifets} 
as a characteristic plasmon energy, where $\hbar\omega_p=28.816\sqrt{Z_i\rho/A_i} [eV]$
(with $\rho$ given in g/cm$^3$).
In our case, this value is $\varepsilon_{min}=0.7\times 10^{-3}$~eV.
$\varepsilon(a)$[eV] is the energy loss given by the mean free path
$l(\varepsilon)$[cm]$=0.823 \varepsilon^{1.156-0.0954\ln(\varepsilon)}/\rho$.

Accounting for beam particle density, the deposited energy becomes:
\beq
\Delta E_i=\Delta E_T N_p I_G.
\label{dEi}
\eeq
The charge emitted by the dust is evaluated by the cross section for beam energy loss
larger than $\varepsilon_{max}$:
\beq
\Delta Q=\frac{\Delta E_Q N_p}{\varepsilon_{max}} I_G. \label{dQ}
\eeq


The temperature rise due to energy deposition  $\Delta E_i$ is  governed  by
the heat capacity $C_T$ and volume: 
\beq
\Delta T=\Delta E_i\left(\frac{3A_i}{4\pi a^3 C_T\rho}\right).\label{dT}
\eeq

Liquefaction occurs when the dust temperature exceeds its melting point. 
If the charge exceeds the threshold, the liquefied dust undergoes fission,
\beq
Q_{th}=\sqrt{\frac{16\pi\gamma_s a^3}{r_e m_ec^2}}, \label{Qth}
\eeq
where $\gamma_s$ is the surface tension.
Regarding the graphite handled here, the liquid state does not exist, 
so there is no need to consider fission.

\begin{figure}[hbt]
  \centering
  \includegraphics[width=60mm]{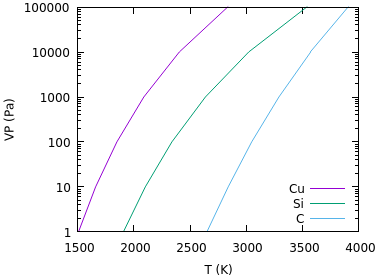}
  \caption{Vapor pressure for Copper, Silicone and Carbon graphite\cite{VPSICO}. }
  \label{VaperP}
\end{figure}

Evaporation becomes significant when temperatures approach sublimation/melting points, 
driven by vapor pressure  $P(T)$. 
Vapor pressures for Cu, Si and Carbon graphite 
are shown in Fig. \ref{VaperP}.

The number of atom $A$ in the dust decreases as:
\beq
\frac{dA}{dt}=-\frac{4\pi c a^2}{\sqrt{2\pi k_B T M_i c^2}}P(T)\label{dAdt}
\eeq
where $M_i=A_i M_{AU}$ is the atomic mass. 

Vaporized molecules are ionized by beam collisions, producing electrons. 
The dust transitions to a plasma state comprising charged dusts, ions, and electrons, 
through which the bunch train propagates while sustaining plasma generation.

For positron beams, the plasma cloud exerts defocusing forces via charged dust particles/ions 
and focusing forces via electrons. 
As these components exhibit distinct spatial distributions, 
their net effect on the beam does not cancel.

\section{Simulation for interaction of beam and dust particles}\label{BDsim}
We perform a numerical simulation that tracks the state of the dust and the motion of beam, 
using the formalisms described in Sec.~\ref{BDpar}. 
The charge distributions of the beam and the dust plasma are represented 
using the Particle-In-Cell (PIC) method in the two-dimensional transverse plane. 
Their respective electric fields are computed by solving the two-dimensional Poisson equation, 
and the motion of particles in the beam (positrons) and the dust plasma (ions and electrons)
is simulated under the influence of these fields.
A positron bunch is represented by 1 million macro-particles. 
The number of macro-ions and macro-electrons fluctuates, but their charges are adjusted to 
maintain a population of approximately 100,000 macro-particles each. 
The dust is modeled as a discrete particle with defined charge and size.

The initial longitudinal extent of the plasma in the simulation is on the order of the dust size. 
At this stage, it is comparable to the beam size, making a purely two-dimensional approximation 
less suitable. However, as the interaction with the beam proceeds, a significant portion of the ions and
 electrons surrounding the beam originate from previous bunches. 
These electrons  with a thermal velocity of $2.1\times 10^5$~m/s spread over $\pm 0.9$~mm
in a bunch interval of 4~ns. Ions and atoms, which move at 1400 m/s ($T=3000$~K), 
spread over $\pm 0.6$~mm in 100 bunch passage (400~ns). 
Since the electrons and ions condensing around the beam are the primary focus, 
the two-dimensional approximation is considered appropriate.

A graphite dust particle with a radius of 50~$\mu$m is placed at the transverse beam center.
Bunches with a charge of $N_p=6.25\times 10^{10}$ pass every $t_{rep}=4$~ns. 
The dust is heated through ionization losses as described by  Eq.(\ref{dET})-(\ref{dEi}) 
and becomes charged via electron emission according to Eq.(\ref{dQ}).
Using Eq.(\ref{dT}), the temperature rise due to ionization loss is calculated to be $\sim 40$~K 
per bunch passage.
The electron energy for which the mean free path equals the dust radius $a=50~\mu$m is
$\varepsilon(a)=79$~keV.
Approximately $1.2\times 10^8$ electrons are emitted per bunch passage according to Eq.(\ref{dQ}), 
simultaneously charging the dust to an equivalent positive charge.

In Ref. \cite{ohmiSBL_eefact25}, 
the interaction between a silicon dust particle and the beam was investigated. 
In the simulation, electrons were emitted continuously without accounting for absorption. 
The dust was heated and charged linearly due to the ionization loss of beam. 
With a surface tension of silicon $\gamma_s=10$~N/m, the threshold of fission 
was calculated as $Q_{th}=5\times 10^8$ from Eq.(\ref{Qth}).
Since the accumulated charge substantially exceeded the threshold at the melting temperature
(T=2000~K), the dust underwent explosive fission around the 80th bunch passage. 
The effect on the beam became noticeable by the 200th bunch and gradually diminished with 
subsequent bunches.

Sudden Beam Loss (SBL) at SuperKEKB, characterized by beam loss occurring over a few turns 
($10 \sim 50$~$\mu$s), exhibits a significantly longer timescale than the 200 bunches 
(0.8 $\mu$s) observed in the simulation. 
The initial pressure bump scenario in Sec.~\ref{VacBump} aligns more closely with 
the observed SBL timescale.

In this study, we consider graphite as the dust material, which undergoes sublimation 
without liquefaction. The simulation was improved to better reflect physical realism by 
incorporating electron emission, absorption, and secondary electron emission. 
Emitted electrons oscillate within the beam potential, with most being reabsorbed upon 
collision with the dust, subsequently triggering secondary electron emission.
Simulations were conducted for dust located in both drift space and under a bending field 
($B_y  =1$~T). A simplified secondary emission model was employed, 
where the secondary emission yield is set to 0.5 for incident electron energies below 50 eV 
and 1.0 for energies $\ge 50$~eV.

The bunch, with a total length of $6\sigma_z=36$~mm is divided into 40-120 slices, 
where $\omega_e\sigma_z/c=9.4$.
The time step for integrating particles motion is $ct_{step}=0.9-0.3$~mm.
The plasma density created by the beam is very high.
In the early stage of the interaction, a dust particle and $\sim 10^8$ electrons are localize 
within a region of $a=50 \mu$m, resulting in a density of $2\times 10^{20}$~m$^{-3}$.
This yields an electron plasma frequency of $\omega_p=\sqrt{4\pi r_e n_ec^2}=2\pi\times 127$~GHz, 
$\omega_p\sigma_z/c=16$.
As shown later the denisty can reach $10^{21}$~m$^{-3}$ as the interaction progresses, resulting in
$\omega_p=2\pi\times 284$~GHz, $\omega_p\sigma_z/c=36$.
The frequency is higher than the electron oscillation frequency in the beam potential, $\omega_e$.

The rapid nature of plasma oscillations makes simulations particularly complex. 
The bunch gap after the beam passage has a length of $ct_{rep}=1.2$~m. 
The motion of electrons and ions during this interval is also of interest. 
However, a preliminary evaluation suggests that plasma oscillations within the bunch gap 
can likely be neglected. 
Therefore, it should be sufficient to focus on interactions between the bunch and the dust plasma 
during their direct interaction. 
Since it is not yet possible to draw definitive conclusions, 
this paper will limit its scope to the interaction between the beam and the dust plasma. 
Interactions within the dust plasma itself, including plasma oscillations, will be addressed in future work.

During the bunch spacing, no beam force is present. In the early stage, 
where only dust and electrons exist, all emitted electrons with energies 
below the dust's potential energy are reabsorbed. 
Secondary electron emission is suppressed during the bunch spacing simulation 
when the dust's potential energy exceeds 10 eV (the characteristic energy for secondary emission).

As the dust temperature approaches or exceeds its sublimation point, evaporation intensifies. 
The dust evaporates as neutral atoms or occasionally as ions, with the rate given by Eq.~(\ref{dAdt}). 
The evaporated atoms are ionized by the beam, and the resulting electrons are absorbed by the dust 
during their movement, potentially inducing secondary electron emission. 
The simulation tracks these processes - generation, motion, and absorption - using macroparticles 
and macrocharges.

\begin{figure}[hbt]
    \centering
    \includegraphics[width=40mm]{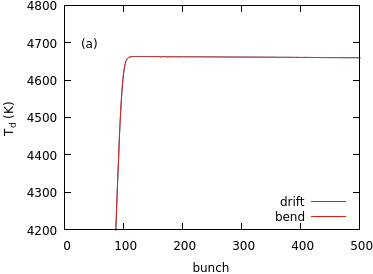}
    \includegraphics[width=40mm]{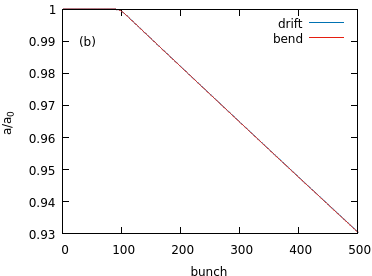}
    \caption{Evolution of (a) dust temperature and (b) dust size.}
    \label{DustTR}
\end{figure}

Figure \ref{DustTR} shows the evolution of the dust temperature and size. 
The dust temperature increases by 40~K per bunch passage and reaches 
the sublimation point (T=3500~K) by the 80th bunch. 
It then rises further to an equilibrium temperature of 4700~K, where the beam-induced 
heating balances the heat of sublimation. The dust size gradually decreases due to evaporation 
as sublimation becomes active, with a reduction rate of $a/a_{0}=1.7\times 10^{-4}$/bunch, 
where $a_{0}=50$~$\mu$m.
The temperature saturates at the 100th bunch.

\begin{figure}[hbt]
  \centering
  \includegraphics[width=40mm]{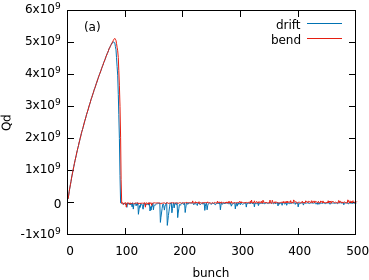}
  \includegraphics[width=40mm]{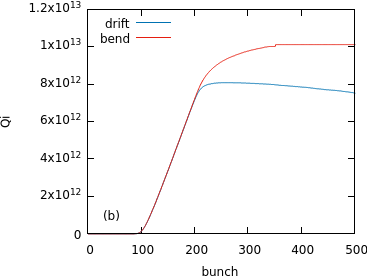}
  \includegraphics[width=40mm]{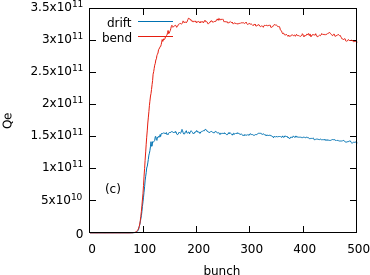}
  \includegraphics[width=40mm]{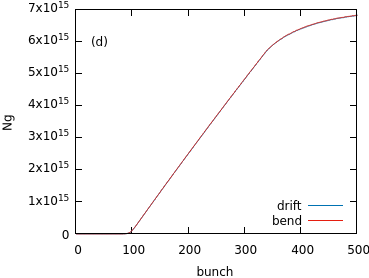}
  \caption{Evolution per bunch passage of  (a) dust charge, (b) ions charge, and 
  (c) electron charge, and (d) the number of evaporated neutral atoms.}
  \label{Qdieg}
\end{figure}

Figure \ref{Qdieg} shows the evolution per bunch passage of (a) dust charge, (b) ion charge, 
(c) electron charge, and (d) the number of evaporated neutral atoms. 
The simulation was run for 500 bunches. 
Electron and ion counts were truncated beyond a radius of 3~cm, while neutral atoms were 
truncated beyond 3~mm. Although charged particles may rapidly move away and occasionally return, 
neutral atoms exhibit only thermal motion and remain slow; once displaced from the beam region, 
they cease to affect the system. By focusing on the vicinity of the beam, 
the system is observed to roughly saturate after 200-300 bunches.
Each bunch passage generates approximately $1\times 10^8$ electrons, leading to dust charging.
The dust charge reaches up to $5\times 10^9$ in both the drift and the bend sections.

After the sublimation point, evaporation becomes significant, 
causing a sharp decline in dust charge. 
This drop to $\sim 10^8$ is primarily driven by the absorption of electrons produced 
through beam ionization of evaporated atoms, rather than direct ionization of the evaporating material. 

In the drift space, the ion and electron charges are in the range of $8\times 10^{12}$ and 
$1.5\times 10^{11}$, respectively. 
Under a bending field, these values change to $1\times 10^{13}$ for ions and $3\times 10^{11}$ 
for electrons. 
The higher electron charge in the bend region is notable. The number of electrons is generally 
lower than that of ions due to their higher mobility.
The number of neutral atoms increases monotonically and is saturated at 400th bunch due to the truncation. 

\begin{figure}[hbt]
  \centering
  \includegraphics[width=50mm]{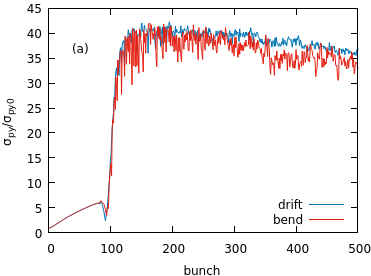}
  \includegraphics[width=36mm]{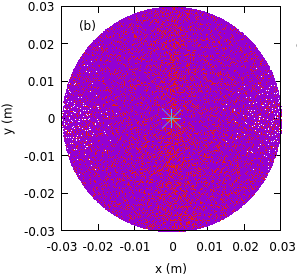}
  \includegraphics[width=44mm]{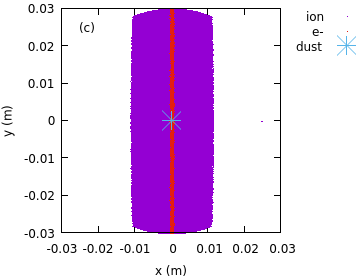}
  \caption{ 
    (a) Evolution of the vertical momentum spread during interaction with dust plasma; 
    (b,c) distribution of the dust plasma in drift space and bending field, respectively. }
  \label{sigpy}
\end{figure}

Positrons in the bunches experience transverse kicks due to the charge distributions of 
the dust, ions, and electrons. While the transverse positions of the positrons remain unchanged, 
their transverse momenta are altered. 
Consequently, the bunch size remains constant, but a momentum spread develops. 
This effect is particularly pronounced in the vertical direction due to the smaller beam size 
in that plane. The change in momentum distribution subsequently induces size variations 
via betatron oscillation.
Figure \ref{sigpy}(a) shows the evolution of the vertical momentum spread. 
The momentum spread is prominent during dust interactions in both drift and bending fields. 
The increase in $\sigma_{py}/\sigma_{py0}\approx 40$ is critical for beam loss. 
Figures \ref{sigpy}(b) and (c) display the macro-particle distributions of dust, ions, 
and electrons after interaction with the 500th bunch. 
The distributions of electrons and ions exhibit distinct features depending on 
the presence or absence of a bending field.

\begin{figure}[hbt]
  \centering
  \includegraphics[width=42mm]{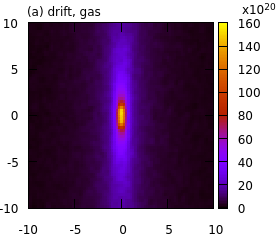}
  \includegraphics[width=42mm]{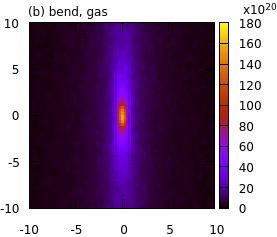}
  \includegraphics[width=41mm]{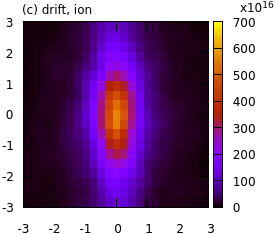}
  \includegraphics[width=41mm]{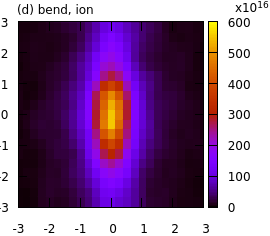}
  \includegraphics[width=41mm]{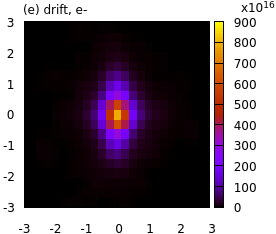}
  \includegraphics[width=41mm]{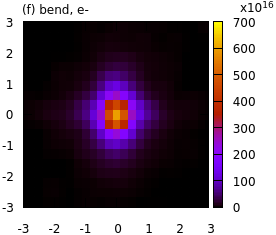}
\caption{Density distributions of dust plasma: (a, b) atomic gas distributions
 in drift and bending fields;
(c, d) ion distributions in drift and bending fields;  (e, f) electron distributions
 in drift and bending fields.
 The color scale represents density in unit of m$^{-2}$. 
The  horizontal and vertical axes represent $x/\sigma_x$ and $y/\sigma_y$, respectively.}
  \label{densityb500}
\end{figure}

Figure \ref{densityb500} shows the density distribution of the dust plasma in the vicinity of the beam. 
The ion and electron distributions represent the density during interaction with the central part
of the bunch  ($z=0$).
The maximum atomic gas density is $\rho_G=1.8\times 10^{22}$~m$^{-2}$, independent of
whether the interaction occurs in a drift or bending section.
This density is 50 times higher than that in the vacuum bump scenario ($\rho_G=5.3\times 10^{20}$~m$^{-2}$) 
in Sec.\ref{VacBump}.
In the case of the vacuum bump, the gas is uniformly distributed, 
whereas here it is localized only in the vicinity of the beam.
The ion and electron densities are largely independent 
of the field configuration, with similar maximum values of $\rho_i=6-7\times 10^{18}$~m$^{-2}$ 
and $\rho_e=7-9\times 10^{18}$~m$^{-2}$, respectively.
It is slightly higher in the drift section.
It can be inferred that the electron distribution varies during interaction 
along the longitudinal bunch position $z$, while the ion distribution remains relatively unchanged.
Consequently, the integrated effect of ions is stronger than that of electrons. 
The observed densities are approximately 30 times higher than those resulting from 
ionization at the vacuum bump ($\rho_i\approx 3\times 10^{17}$~m$^{-2}$), as discussed in Sec.\ref{VacBump}.

\begin{figure}[hbt]
  \centering
  \includegraphics[width=41mm]{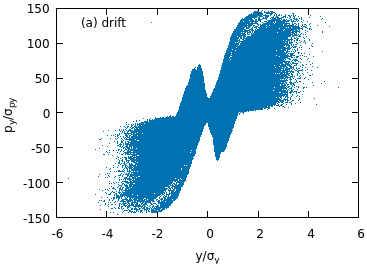}
  \includegraphics[width=41mm]{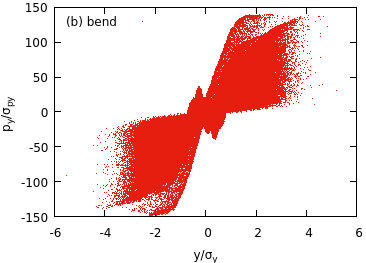}
  \includegraphics[width=41mm]{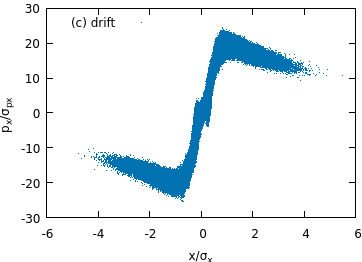}
  \includegraphics[width=41mm]{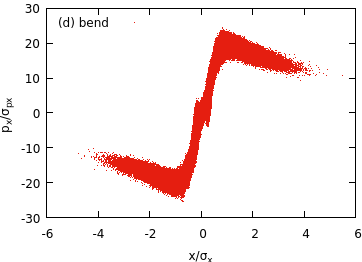}
  \caption{Phase space distribution of 500th bunch: 
  (a) $(y,p_y)$ plot for interaction in drift space,  (b) $(y,p_y)$ plot for interaction in a bending field,
  (c) $(x,p_x)$ plot for interaction in drift space,  (d) $(x,p_x)$ plot for interaction in a bending field. }
  \label{ypyDB}
\end{figure}

Both electrons and ions are distributed near the beam, but their densities and spatial profiles differ.
Particles in the positron bunch experience a defocusing force from the ion distribution 
and a focusing force from the electron distribution. 
The net momentum kick resulting from this mixture of focusing and defocusing forces induces 
a transverse momentum spread $\sigma_{py}$ within the bunch. Note that the beam size $\sigma_y$
remains unchanged at the instant of interaction.
Figure \ref{ypyDB} shows the phase space distribution of 500th positron bunch. 
Plots (a) and (b) correspond to the vertical phase space distribution under 
the drift and bending field configurations, respectively. 
Plots (c) and (d) show the horizontal phase space distribution. 
These plots exhibit predominantly defocusing forces overall, 
but a focusing component is also observed in regions where the amplitude is small.
This behavior is consistent with the density distribution shown in Fig. \ref{densityb500}.
The vertical and horizontal distributions differ significantly. 
The gas distribution is approximately the same size as the beam. 
Ions and electrons, created within the overlap region of the beam and gas, 
are generated with a size smaller than that of the beam, 
and the horizontal size is larger than the vertical one. As a result, $dp_y$
depends on $x$ and exhibits significant spreading, whereas $dp_x$, being independent of $y$, 
shows less spreading.

\section{Multi-turn simulation}\label{MultiTurn}
In SuperKEKB, each ring stores 2346 bunches. 
Simulating the interaction of all these bunches with dust over multiple turns is 
computationally difficult. Therefore, a simplified approach, described below, is adopted.

When the interaction with dust causes it to reach the sublimation point following the initial 
temperature rise, the resulting dust plasma interacting with the bunch reaches 
a steady-state distribution. 
As shown in Fig.~\ref{sigpy}(a), this steady state is achieved within 100 bunch passages 
after sublimation occurs. Although the momentum spread is observed to gradually decrease thereafter, 
assuming it remains constant at a fixed value enables a simplified multi-turn simulation.

The momentum change of the bunch induced by the dust plasma remains consistent each time, 
resulting in an identical phase-space distribution.
By applying the revolution matrix, the bunch distribution after one turn can be obtained. 
Repeating the interaction between this bunch and the dust plasma yields 
a new steady-state dust plasma distribution, while simultaneously producing 
a corresponding steady-state phase-space distribution of the bunch interacting with it. 
This process is iterated over multiple turns. 
Thus, by simulating the interaction of a single bunch with the dust plasma, 
it is possible to effectively model multi-bunch and multi-turn dynamics.
The beam-dust interaction over 10 turns is simulated through a total of 
500+100$\times 10$=1500 times interactions, where 
the bunch was made to interact (with the dust) 100 times to generate a steady-state plasma 
distribution each revolution.

In the simulations presented in this section, to make beam loss more visible, 
the minimum energy deposition threshold for dust was raised from 
$\varepsilon_{min}=0.7\times 10^{-3}$~eV to 10~eV. 
As a result, the stopping factor in Eq.(\ref{dET}) is reduced by $\log(10/0.7\times 10^{-3})=9.6$.

\begin{figure}[hbt]
  \centering
  \includegraphics[width=40mm]{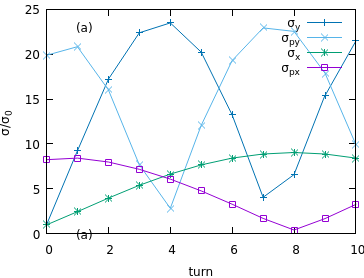}
  \includegraphics[width=40mm]{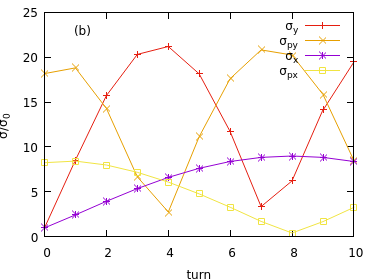}
  \caption{Evolution of the vertical beam size and momentum spread:
(a)  Vertical beam size and momentum spread  in drift space, 
(b) Vertical beam size and momentum spread in bending field. }
 \label{TbTsigyBDr10}
\end{figure}

Figure \ref{TbTsigyBDr10} shows the turn-by-turn evolution of the beam size. 
The vertical and horizontal beam size and momentum spread 
are plotted in (a) and (b) for the drift and bending sections, respectively.
The beam size in the $y$-direction increases at the 4th and 10th turns, 
while the size in the $x$-direction increases at the 8th turn. 
This behavior originates from the tune values of $(\nu_x,\nu_y)=(0.53,0.57)$. 
At other locations in the ring, the turn at which the size becomes large varies 
depending on the betatron phase advance from the point of dust interaction.
Considering both the beam size and momentum spread, the emittance remains nearly constant 
from the first interaction in both the drift and bending sections. 

\begin{figure}[hbt]
  \centering
  \includegraphics[width=40mm]{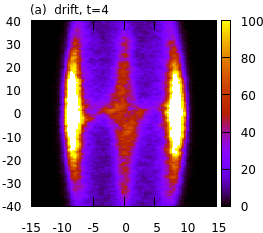}
  \includegraphics[width=40mm]{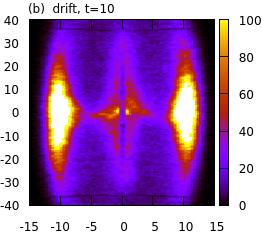}
  \includegraphics[width=40mm]{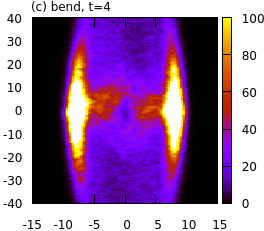}
  \includegraphics[width=40mm]{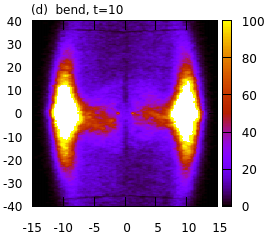}
  \caption{Particle distribution in a bunch after dust interactions:
(a) 4th turn  (b) 10th turn  in drift space, 
(c) 4th turn, (d) 10th turn in a bending field.
 The  horizontal and vertical axes represent $x/\sigma_x$ and $y/\sigma_y$.
 The color scale represents number distribution of macro-particles in a bunch, 
 where the total number is $10^6$.}
  \label{xydistt3-10}
\end{figure}

Figure \ref{xydistt3-10} shows particle distribution of the bunch after turns at the dust position.
The distributions after 4th and 10th turn in drift space are plotted in (a) and (b).
The distributions in bending field are plotted in (c) and (d). 
The distribution in the x-direction exhibits a highly distinctive feature where the beam splits 
into two distinct profiles at certain turns. 
This can be explained by the phase space distribution shown in Fig. \ref{ypyDB}. 
Since $\sigma_x\gg\sigma_y$, there is no spread in the $x$ phase space distribution, and
since the plasma distribution is small, even the core part of the beam experiences nonlinear kicks. 
Nevertheless, the key point is that the overall distribution spreads significantly in the $y$-direction, 
and beam loss is ultimately determined by the vertical distribution.

The PIC grid space is set to a maximum of $40\sigma$ in the $y$-direction, matching the aperture. 
Particles that exceed this grid space account for 10\% in both the drift and bend sections 
at the 4th turn.
To analyze beam loss accurately, it is necessary to track the particle distribution in phase space 
through the accelerator using a beam tracking code and examine the losses at the collimator. 
Since detailed information such as collimator settings is required, 
this analysis will likely be conducted separately.

\section{Conclusion}\label{conclusion}
We have been investigating the effects of beam-dust interactions on the positron beam. 
Graphite particles with a diameter of 50~$\mu$m is assumed as the dust. 
The beam loses energy due to ionization loss in the dust. 
The dust, in turn, loses electrons and becomes charged, while the energy received from 
the beam causes it to heat up, evaporate, and generate atomic gas. 
This gas is ionized by the beam, resulting in the formation of plasma around the beam. 
Unlike conventional neutral plasma, this plasma consists of dust, ions, and electrons, each 
with distinct charge distributions. 
These charge distributions exert focusing and defocusing forces on the beam, 
instantaneously altering the phase-space distribution of the beam particles.

Simulations based on this scenario were conducted, tracking the distributions of dust, 
atomic gas, ions, and electrons after each bunch passage, and calculating the resulting 
momentum changes experienced by the bunch. The results indicate that particles within 
the bunch can exhibit betatron amplitudes of up to approximately 40$\sigma_y$ 
(the aperture limit), suggesting the possibility of instantaneous beam loss. 
This concept is considered highly likely to explain the Sudden Beam Loss (SBL) observed in SuperKEKB.

The density of the dust plasma is higher than that of the beam. 
As a result, the electron plasma oscillations within the dust plasma occur faster than 
the electron oscillations induced by the beam, making it necessary to use a finer time step 
and perform careful calculations for the space charge within the dust plasma. 
Although this paper does not address plasma oscillations inside the dust plasma, 
we intend to investigate them in a future study.

Similarly, the same simulation can be applied to electron beams; however, 
the internal state of the resulting plasma is expected to differ significantly. 
We intend to discuss this in a future study.


\section{acknowledgement}
The work is supported by the Chinese Academy of Sciences President's International Fellowship 
Initiative (Grant No. 2024PVA0057) and innovation Study of IHEP. 
The authors thank T.~Abe, Y.~Funakoshi, H.~Ikeda, M.~Kikuchi, A.~Lechner, G.~Mitsuka, T. ~Miura, 
G.~Nigrelli, M.~Sallivan, Y.~Zhang and F. Zimmermann
for their fruitful discussions.

\end{document}